\documentclass[nofootinbib,reprint,superscriptaddress,preprintnumbers,amsmath,amssymb,amsfonts,aps,prl,floatfix]{revtex4-2}

\usepackage{xcolor}
\usepackage{hyperref}
\usepackage{graphicx}
\usepackage{dcolumn}
\usepackage{bm}
\usepackage{float}
\usepackage{verbatim} 
\usepackage{amsmath}
\usepackage{amssymb}
\usepackage[dvipsnames]{xcolor} 
\usepackage{amssymb}
\usepackage{comment}
\newcommand\david[1]{{\textcolor{blue}{DL: \bf #1}}}

\begin{document}

\preprint{UdeM-GPP-TH-26-311}

\title{Searching for New Physics with Reinforcement Learning}

\author{Jacky Kumar}
\email{jacky.kumar@umontreal.ca}
\affiliation{Physique des Particules, Universit\'e de Montr\'eal, 1375 Avenue Th\'er\`ese-Lavoie-Roux, \\ Montr\'eal, QC H2V 0B3, Canada }

\author{Marianne Bouchard}
\email{marianne.bouchard.5@umontreal.ca}
\affiliation{Physique des Particules, Universit\'e de Montr\'eal, 1375 Avenue Th\'er\`ese-Lavoie-Roux, \\ Montr\'eal, QC H2V 0B3, Canada }

\author{David London}
\email{london@lps.umontreal.ca}
\affiliation{Physique des Particules, Universit\'e de Montr\'eal, 1375 Avenue Th\'er\`ese-Lavoie-Roux, \\ Montr\'eal, QC H2V 0B3, Canada }

\date{\today}

\begin{abstract}
Finding new physics (NP) is the fundamental problem in particle physics today. Studying ``anomalies,'' i.e., measurements of low-energy observables whose values disagree with the predictions of the Standard Model (SM), is a powerful search strategy. The SM Effective Field Theory (SMEFT) provides a general model-independent framework for parametrizing NP; given some anomalies, it is natural to try to find the SMEFT operator(s) that can explain them. This is a challenging task because (i) the number of SMEFT operators is enormous, and (ii) at loop level there are very complicated correlations among the operators. Analyses by humans typically rely on phenomenological intuition to decide which operators are relevant. This is often biased and does not explore the complete SMEFT operator space. Interestingly, reinforcement learning (RL) techniques excel at tasks that require decision making to achieve their goals. In this paper, we introduce an RL method that can be used to find the SMEFT operators that explain any anomalies. 
Using some explicit examples, we demonstrate that our RL method can be used to efficiently search for NP at the level of SMEFT. 
The RL training can be performed efficiently using standard CPU resources, without requiring specialized computing hardware.
\end{abstract}


\maketitle

As artificial intelligence (AI) has grown in importance, machine learning (ML) has been used to address problems in a wide variety of areas. 
Naturally, this raises a compelling question: can AI systems be leveraged to solve fundamental problems in theoretical particle physics? To date, there have been only a few attempts to apply ML to such problems \cite{Alexander:2026lpw, Baretz:2025zsv, Hammad:2026bvw, Saad:2026pan}.  In this paper, we demonstrate that ML can help in tackling the fundamental open problem in particle physics today, namely the search for physics beyond the Standard Model (SM).

To address this problem, we use a type of ML approach known as reinforcement learning (RL), which is best suited for tasks that can be formulated as sequences of decision making with the aim of optimizing some objective \cite{sutton2018reinforcement}.
In RL problems, the goal is to find a \emph{policy} that maps \emph{states} to 
\emph{actions} (i.e., a decision-making rule). The policy learns through repeated interactions with an \emph{environment} 
using \emph{rewards} as a feedback signal. 
Here, ``state'' refers to the configuration of the environment at a given time. 
To be specific, solving an RL problem amounts to finding a policy that has learned a
probability distribution over actions that assigns a high (low) probability to actions leading to a high (low)  long-term reward. Below, we describe the physical problem that we wish to solve. Afterwards, we detail how the general RL approach connects to this problem.

The SM of particle physics has been enormously successful in describing the physics up to energy scales of $O({\rm TeV})$. Even so, it is not complete: it cannot explain a number of observations, such as neutrino masses, dark matter, the baryon asymmetry of the universe, etc. We therefore conclude that there must be physics beyond the SM.

As the large hadron collider has not found any new particles, this new physics (NP), whatever it is, must be heavy. The effects of such NP can only be seen indirectly. That is, the virtual exchange of NP particles will affect certain low-energy processes. This will show up as an ``anomaly:'' the measurements of observables associated with an affected process will disagree with the predictions of the SM.

The approach most appropriate for analyzing anomalies involves model-independent effective field theories (EFTs). At high energies, when the NP is integrated out, one obtains the Standard Model EFT, SMEFT \cite{Buchmuller:1985jz, Grzadkowski:2010es, Brivio:2017vri, Aebischer:2025qhh}, which contains only SM particles. In SMEFT, the leading-order (dimension-4) terms are those of the SM.
The NP produces contributions to the higher-order terms, which are suppressed by powers of the NP scale $\Lambda$.

Whenever an anomaly is reported, the obvious question is: what type(s) of NP could explain it?
In order to answer this, it is useful to perform a bottom-up analysis, which consists of finding which SMEFT operators can best explain the anomalies. 
In Ref.~\cite{Grzadkowski:2010es}, all dimension-6 operators were tabulated. (This is known as the Warsaw basis.) 
In our study, we focus only on those operators that conserve baryon and lepton number, and also lepton flavour. There are a total of 912 such operators when individual flavour indices are considered.

Humans typically use their phenomenological intuition to guide their search for these SMEFT operators. This intuition requires an understanding of many different issues: the structure of the SMEFT operators, how to interpret experimental data, and the relationship between the operators and the data. 

If there is only a single anomaly, this intuition works reasonably well. But things start to get complicated as the number of anomalies increases. Because the total number of SMEFT operators that contribute to a given set of observables can be large, and because these operators are correlated in non-trivial ways, this leads to a huge combinatorial search space. In this case, humans will typically choose SMEFT operators based on their favourite NP theory, or simply by trial and error.

In this paper, we formulate this task as an RL problem.
We begin by describing the general physics problem.
First, a NP SMEFT operator is considered. In order to see how well this operator explains the experimental data, a $\chi^2$ test is performed using the Wilson coefficient (WC) of the operator as an unknown parameter. For this purpose, the program \texttt{flavio} \cite{Straub:2018kue} is used to calculate the theoretical predictions for the observables as a function of the WC, and these are used to compute the $\chi^2$ function. The value of the WC that minimizes the $\chi^2$ is found. 
The general goal is to find a minimal subset of operators that most reduces the $\chi^2$.

With this description, we can make the connection between the elements of the RL method and the physics problem. The correspondence is as follows: 

\begin{itemize}

\item {\bf Policy:} the decision-making rule used to propose SMEFT operators,

\item{\bf Action:} proposing an operator from the 912 SMEFT operators used in this study (this set of operators is known as the vocabulary),

\item{\bf Environment:} the observable calculator, the $\chi^2$ function, and the $\chi^2$ minimizer,

\item{\bf State:} the set of proposed SMEFT operators and the value of $\chi^2_{\rm min}$,

\item{\bf Reward:} the improvement in the global fit, i.e. the reduction in $\chi^2$, when adding the proposed operator.
\end{itemize}

Our RL problem is therefore to find a policy to  propose operators that can best explain the data. The policy is trained by repeatedly sampling the proposed operators from it, receiving the resulting rewards from the environment, and updating it to improve its performance. 
Over time, the policy learns to propose operators that yield the largest decrease in the value of $\chi^2_{\rm min}$, i.e., the largest reward. In what follows, we describe the essential technical aspects of this RL framework.

During training, sets of $T$ operators, called trajectories, are constructed iteratively. At each step $t$, a new operator is proposed. The operator is sampled from the policy $\pi_\theta(s_t)$, which gives a probability distribution over the vocabulary for a given state $s_t$. For instance, $\pi_\theta(a_t|s_t)$ gives the probability that an action $a_t$, i.e. an operator $O_t$ that is not already in the state, is chosen given state $s_t$.
The state is then updated to include the previous proposed SMEFT operators (from state $s_{t-1}$) and the new operator, as well as the new value of $\chi^2_{min}$, obtained by fitting the WCs of the selected operators. 

A reward $r_t= \chi^2_{min,t} - \chi^2_{min,t-1}$ is assigned at each step $t$, and represents the improvement in $\chi^2_{min}$ following the addition of the proposed operator. This reward indicates the usefulness of an operator relative to others, and is used to update the policy.

The policy is parametrized by a transformer-based neural network (NN) \cite{sutton1999policy,vaswani2023attentionneed}. The training proceeds by iteratively sampling trajectories from the current policy $\pi_\theta$ and using them to update the NN's parameters by minimizing the loss with AdamW (parameters are updated once per batch, which contains 25 rollouts, i.e., trajectories). 

The loss is defined as
\begin{equation} \label{eq:actorloss}
\mathcal{L}(\theta) = -{1\over T} \sum_{t=1}^{T} \hat A_t \log \pi_\theta (a_t|s_t) ~.
\end{equation}
The expected return $A_t$ is normalized for each step over the batch to get $\hat A_t$. The expected returns are computed by the sum of the step rewards of the remaining steps:
\begin{equation}
    A_t = r_t+ r_{t+1} + ... + r_{T}~.
\end{equation}

The training progressively updates the probability distribution over the vocabulary, i.e., the policy, to increase the likelihood of choosing operators that most reduce the $\chi^2$.

The algorithm described above is a simplified version of the full actor--critic framework implemented in the GitHub repository \href{https://github.com/jackypheno/ReinforcePhysics}{ReinforcePhysics}. For the numerical analysis presented in this work, we use various codes available in this repository and adapt them to the specific requirements of our study. Our simplified code can be found on \href{https://github.com/Marianne88/ReinforcePhysics/blob/Marianne}{GitHub}.

We now provide two examples demonstrating the application of this RL method to the search for NP. First, we consider only a single anomaly, the CDF $W$-mass anomaly \cite{CDF:2022hxs}. This was studied in Ref.~\cite{Bagnaschi:2022whn}, and sets of NP SMEFT operators were found that can explain the anomaly. We show that our RL method is able to successfully reproduce the results of Ref.~\cite{Bagnaschi:2022whn}. Importantly, this method gives much more information. Rather than simply identifying the SMEFT operators that best explain the data, the RL method finds all sets of SMEFT operators that produce acceptable fits to the data. Given that we would like to identify all possible NP explanations, this is extremely useful.

In the second example, in order to test the method in a more challenging scenario, we create a synthetic dataset with multiple fictitious anomalies. This example better exploits the capabilities of the RL algorithm, since multiple correlations among operators and observables must be taken into account in order to solve this problem.

We begin with the first example. Several years ago, the CDF Collaboration measured the mass of the $W$ boson, finding $m_W= 80433.5 \pm 9.4 ~{\rm MeV}$ \cite{CDF:2022hxs}.
There is a significant tension between this value and the SM prediction obtained from precision electroweak data, $m_W= 80354 \pm 7 ~{\rm MeV}$ \cite{Haller:2018nnx},
as well as with previous direct measurements. If we take the CDF measurement at face value, the question is: which SMEFT operators can be added to the SM Lagrangian to account for it? This can be addressed using our RL method.

SMEFT operators contribute to many observables. Therefore, when adding SMEFT operators to explain an anomaly, one must also include in the analysis the measurements of all the observables to which the SMEFT operators contribute. These will provide additional constraints on the WCs. 

For the analysis of the $W$-mass anomaly, we use the world average of $m_W$ from Ref.~\cite{Bagnaschi:2022whn}, which includes the anomalous CDF measurement: 
\begin{equation}
    m_W= 80411 \pm 8 ~{\rm MeV} ~.
\end{equation}

For the additional constraints, we include the $Z$-pole observables used in Ref.~\cite{Bagnaschi:2022whn}, and add other electroweak precision observables for a total of 26 \cite{ALEPH:2013dgf, ALEPH:2005ab, ParticleDataGroup:2024cfk} (details can be found in the code, see \href{https://github.com/Marianne88/ReinforcePhysics/blob/Marianne}{github}). 
The measured values of all observables agree with the predictions of the SM, with one exception: the measurement of the forward-backward asymmetry in $Z^0 \to b {\bar b}$, $A_{\mathrm{FB}}^{0,b}$, disagrees with the SM prediction by $2.7\sigma$. In other words, its inclusion in the analysis introduces another anomaly. We (and Ref.~\cite{Bagnaschi:2022whn}) are therefore searching for sets of SMEFT operators that simultaneously explain both anomalies.

Within the SM, the fit yields $\chi^2_{\rm SM}=51.8$. This is the starting point for our RL method. (Note that this value is computed using the default \texttt{flavio} precision, which introduces an error of $\sim7\%$ in the theoretical uncertainties, and can lead to a numerical fluctuation in the value of $\chi^2_{\rm SM}$. However, this does not affect the method itself.)

We ran the RL algorithm for $200$ batches, with 25 trajectory rollouts per batch and $T=10$ time steps per trajectory. This gave us many sets of $10$ operators. However, with only two anomalies, a set of 10 operators will likely include some that contribute negligibly to the reduction of $\chi^2$. We found that the minimal number of operators required to get an acceptable fit (defined below) is two. To identify the minimal subsets of operators that can explain the data, we retain the pairs of operators with the highest drop in $\chi^2$ among the high-reward trajectories identified by the ML algorithm.

The pair of SMEFT operators with the highest reward identified by the policy is (operator definitions can be found in Ref.~\cite{Grzadkowski:2010es})
\begin{equation}
\begin{aligned}
&[O_{ll}]_{1221} =\left(\bar{l}_1\gamma_\mu l_2\right)
\left(\bar{l}_2\gamma^\mu l_1\right) ~, \\
&[O_{HD}] =  \left(H^\dagger D_\mu H\right)^\ast
\left(H^\dagger D^\mu H\right) ~,
\end{aligned}\label{eq:best_pair}
\end{equation}
with $\Delta \chi^2=24.2$. This corresponds to $\chi^2_{\rm min}/{\rm d.o.f.}=1.10$ (p-value of $33\%$), showing that the data is well explained by the addition of these operators. 
Specifically, at the point of minimum $\chi^2$, we obtain $m_W= 80411$ MeV, which matches the measured world average.
This solution agrees with the result of Ref.~\cite{Bagnaschi:2022whn}, which was obtained without the use of  ML.

However, from the point of view of finding NP explanations of anomalies, we are not interested only in the solution with the smallest $\chi^2_{min}$ (i.e., the largest $\Delta \chi^2$). On the contrary, we are interested in all solutions with the minimal number of operators that provide an acceptable fit to the data.
For illustrative purposes, we choose the definition of acceptable as any pair of operators that gives a fit with a p-value $\ge10\%$. 
The policy identifies $77$ such pairs. (This large number of solutions comes from the conservative definition of an acceptable fit. Had we chosen to define an acceptable fit as one with a p-value $\ge20\%$, this number would be considerably smaller.) 

There are several things that we learn from this example. First, the analysis of Ref.~\cite{Bagnaschi:2022whn} is based on human phenomenological intuition, and it was able to correctly identify the pair of SMEFT operators that best explains the two anomalies. Second, not only does the RL method reproduce this result, it identifies many other acceptable solutions. This shows that, even with a small number of anomalies, the RL method provides important new information.

We now turn to testing our method in the context of a more challenging scenario. To this end, we construct ``fictitious'' anomalies involving observables in processes with $\Delta F=1$ and $\Delta F=2$, as well as flavour-conserving transitions. This diverse set of processes probes different classes of SMEFT operators with non-trivial correlations involving a large portion of SMEFT operator space. The resulting setup thus provides a more complex environment and allows us to test the scalability and search capabilities of our RL method. 

We assign fictitious, i.e., invented, values of the measurements to ten observables \footnote{Note that, although we present this complicated scenario of ten anomalies to illustrate certain properties of our RL method, this scenario is unlikely to ever occur in practice. After all, there will be an announcement for each individual anomaly, and after this announcement there will surely be studies (perhaps using the RL method) looking for common NP explanations for all the anomalies to date. So by the time we reach ten anomalies, we will already have a reasonable idea of the types of NP that can simultaneously explain smaller subsets of anomalies.}. The list of observables is given in Table \ref{tab:obs_fictitious}.
These observables are chosen such that they deviate from the SM predictions at the level of 3-5$\sigma$.

We include some additional observables to provide constraints on the selected operators. These extra observables, whose values are taken to be the experimental measurements, are mostly taken from the $Z$-pole observables. Further details can be found in the code available on \href{https://github.com/Marianne88/ReinforcePhysics/blob/Marianne}{github}.
Our dataset now includes 32 observables, the ten fictitious anomalies in Table \ref{tab:obs_fictitious} and 22 others.

\begin{table*}[htb]
\begin{tabular}{|c|c|c|}
\hline \hline 
Observable &  fictitious experimental value & fictitious deviation from SM\\ \hline 

$A_{\mathrm{FB}}^{0,b}$  & $(10.24 \pm 0.03) \times 10^{-2}$ & $-3.2\sigma$ \\

$m_W$  & $80.399 \pm 0.008$  &  $+3.7\sigma$\\

$\mathcal{B}(B^+\to K^+\nu\bar\nu)$  & $(0.60\pm0.02)\times 10^{-5}$ & $+4.1\sigma$  \\

$\frac{d\mathcal{B}}{dq^2}\left(B_s\to\phi\mu^+\mu^-\right)|_{q^2\in[1,6]}$ & $(2.5\pm0.6)\times10^{-8}$ & $-3.4\sigma$  \\
$R_D\equiv
\frac{\mathcal{B}(B\to D\tau\bar{\nu})}
{\mathcal{B}(B\to D\ell\bar{\nu})}\,, ~ \ell=(e,\mu)$ &  $0.328\pm0.007$ & $+3.5\sigma$  \\
$P_5'\left(B^0\to K^{*0}\mu^+\mu^-\right)\big|_{q^2\in[4,6]}$ & $-1.16\pm0.07$ & $-3.9\sigma$  \\
$\mathcal{B}(K_L\to\pi^0\nu\bar\nu)$ & $(4.3\pm0.3)\times10^{-11}$ & $+3.6\sigma$  \\
$\mathcal{B}(K^+\to\pi^+\nu\bar\nu)$  &  $(11.2\pm0.05)\times10^{-11}$ & $+4.3\sigma$ \\
$\varepsilon'/\varepsilon$ & $(-36\pm6)\times 10^{-4}$ & $-4.5\sigma$  \\

$\Delta M_s$ & $(8.1\pm0.5)\times10^{-12}$ & $-4.8\sigma$  \\
\hline \hline
\end{tabular}
\caption{The ten fictitious anomalies. The fictitious experimental values and deviation from the SM are given. }
\label{tab:obs_fictitious}
\end{table*}

\begin{figure}[h]
    \centering
    \includegraphics[width=\columnwidth]{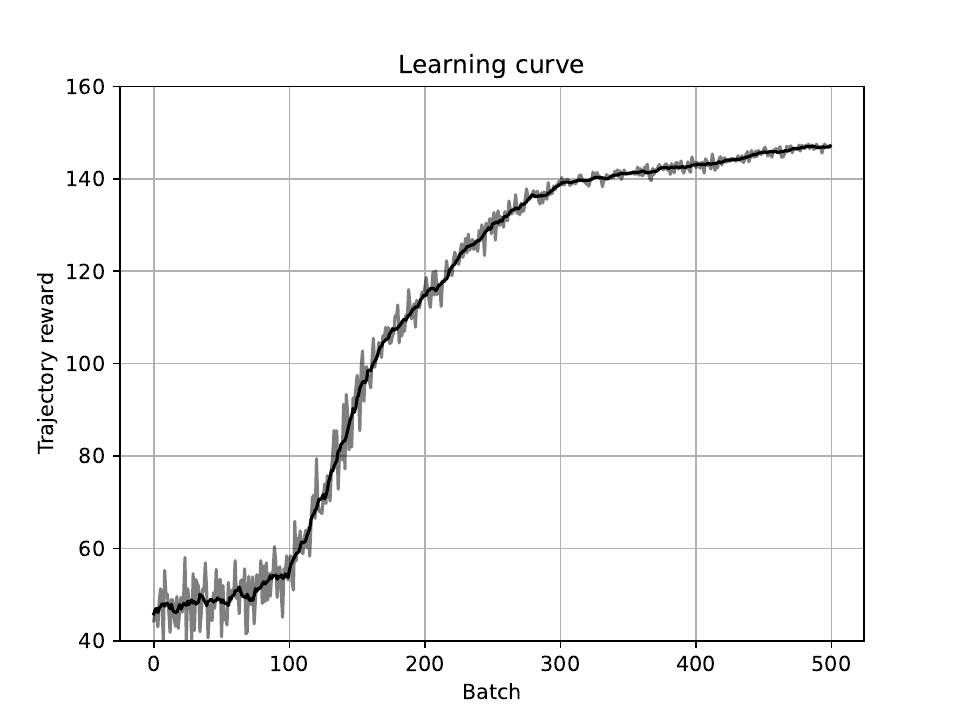}
    \caption{Learning curve over 500 training batches for the fictitious case. The figure shows the mean trajectory reward per batch. The trajectory reward corresponds to the sum of the step rewards for a trajectory (the raw values are shown in faded and smoothed shown in solid are averaged over 10 rolling batches.)}
    \label{fig:learning-curve}
\end{figure}

The SM fit has $\chi^2_{\rm SM}=178.3$.
We train the policy for 500 batches, sampling SMEFT operators with trajectory length $T=10$ and 25 trajectories per batch. Fig.~\ref{fig:learning-curve} shows the evolution of the average of trajectory rewards over the 25 rollouts in a batch. A moving average over 10 batches is also shown. The trajectory reward is defined as the sum of the step rewards, which corresponds to the drop in $\chi^2$ resulting from the addition of the 10 operators in the trajectory. This training takes roughly 30 minutes on a Dell XPS 15, 16 CPUs, Intel i-5.

In the figure, we see an exploratory phase at the beginning during which the reward does not increase much (batches $\sim0$-$100$), followed by a rapid learning phase, in which the policy converges towards operators that give high rewards (batches $\sim100$-$300$). After this, the learning slows down until batch $\sim450$ when the learning comes to a stop.

The analysis of the results is similar to that done in the $m_W$ case: we identify the minimal number of operators required to get an acceptable solution (p-value$\ge10\%$), and we count the number of such subsets found by the RL algorithm. 
It was found that the minimal solutions should have 6 operators to achieve a p-value $\ge 10\%$.
The RL algorithm finds 21 unique subsets that respect these criteria. 

Different baselines can be used to compare this result to non-ML techniques. First, we can compare this to a random search, where sets of 10 operators are chosen from the vocabulary and we evaluate subsets of 6 operators that have p-value $\ge10\%$. When random search evaluates as many sets of 10 operators as the RL algorithm ($N_{\rm eval}= 500\times 25=12500$), no solution is found with a p-value $\ge10\%$. The best subset of 6 operators found has a p-value of $0.04\%$. 

Another baseline we can use is a genetic algorithm \cite{genetic:1989}. This non-RL method, inspired by evolution, creates a population of multiple operator sets and evolves it over many generations. The next generation is constructed by gene selection of members of the current population and random mutation. Population members with good genes, i.e., a high reward, are more likely to pass their genes (operators) to the next generation. Over the generations, only the best operators are kept, while inferior operators will naturally disappear from the population. Once again, we evaluate subsets of 6 operators for high-reward sets and find, with $12500$ evaluations, 16 minimal subsets with a p-value$\ge10\%$. With these baselines, we see that our RL algorithm shows an improvement over these non-RL methods.

In this paper, we have constructed an algorithm, based on reinforcement learning, that can identify the (sets of) SMEFT operators that can explain experimental anomalies. This shows that machine learning can indeed help to address fundamental problems in theoretical particle physics. The method is particularly useful when the space of possible solutions is very large, i.e., when many anomalies are involved, as traditional methods have serious difficulties negotiating this space. On the other hand, even when the number of anomalies is small, we find that the algorithm provides very useful information that non-RL methods don't always give. The point is that this RL algorithm can be thought of as a new technique for finding NP explanations of anomalies, and it can be used regardless of the number (or type) of anomalies.

\acknowledgments
 This work was financially supported by the Natural Sciences and Engineering Research Council of Canada (NSERC) and by FRQNT, Scholarship No.\ 363240 (M.B.).  This research was enabled in part by support provided the Digital Research Alliance of Canada \url{(www.alliancecan.ca)}.  Computations were performed on the Trillium supercomputer at the SciNet \cite{Loken_2010} HPC Consortium and on the Rorqual supercomputer, provided by Calcul Québec. SciNet is funded by Innovation, Science and Economic Development Canada; the Digital Research Alliance of Canada; the Ontario Research Fund: Research Excellence; and the University of Toronto.

\bibliography{NP_with_RL}
\end{document}